\documentclass[prb,preprint]{revtex4-1} % style for Physical Review B and AJP are similar

\usepackage{amsmath} % need for subequations
\usepackage{amssymb}

\usepackage{graphicx} % for figures
\usepackage{epstopdf}
\usepackage{enumerate}
\begin{document}

\raggedbottom

\title{A Wave-centric View of Special Relativity}
\author{William M. Nelson}
\email{wmn@cox.net} %optional
\date{\today}

\begin{abstract}

An approach to special relativity is outlined which emphasizes
the wave and field mechanisms which physically produce the relativistic effects, with
the goal of making them seem more natural to students by connecting more explicitly
with prior studies of waves and oscillators. 

 \end{abstract}
\maketitle

%\tableofcontents

\section{Introduction}

Special Relativity developed simultaneously with the first fundamental theory based on fields and waves, namely 
electromagnetism, but it is usually taught to students in the context
of particle kinematics, using an axiomatic approach in which spacetime is implicitly presented 
as a causative agent.  This can be quite confusing
as it seems to place relativity on a different footing from other physical 
theories, even theories of symmetry principles. 

Rotational symmetry, for example, 
is not typically introduced in terms of axioms or properties of spacetime, but rather in terms
of concrete properties of force laws and potentials. Non-symmetric examples
are readily produced, allowing the student to understand the content of the symmetry
principle. With special relativity, by contrast, one
typically only discusses the two cases of Lorentzian and Galilean symmetry, attributing
both to inherent properties of spacetime, and students may not even realize that there
exists a vast range of easily constructible theories which do not satisfy either principle. 

Approaches to relativity which place a greater emphasis on mechanism have been
discussed periodically over the years, and lately seem to be gaining some
momentum under the rubric of ``Lorentzian''
or ``constructive'' approaches.\cite{bell,brown,miller,rmr} 
In the following we attempt to advance this program by tracing the major relativistic effects
as directly as possible to their origins, which lie most often in the elementary 
behavior of waves or oscillators.  

Many relativistic phenomena which seem strange or opaque in the context of particle kinematics, or
in an axiomatic derivation, are quite natural in the context of the field-
and wave-based theories which actually appear to underlie nature. Indeed they
are often connected to intuitions so commonplace that it is a bit difficult to accept
that they could be relevant to recondite matters such as the ``flow of time''; yet, 
any physical effect must come from somewhere, and the more generic the effect
the simpler the underlying cause ought to be. 

Incorporating such material into standard courses would, we believe, 
have at least the following benefits:

\begin{enumerate}
\item Allow students to build from things they already understand, e.g. 
transverse waves on a string, rather than being confronted immediately with abstract principles.
\item Provide a better historical context for the theory, connecting with partial results
which were obtained by Lorentz, Poincar\`{e} and others before Einstein's formulation.\cite{bell, brown} 
\item Provide tools for reasoning about non-Lorentz-invariant theories, which have been
increasingly under discussion.\cite{kost,kost2} 
\item Provide better preparation for the deeper study of relativistic quantum mechanics or quantum field theories. 
\end{enumerate}

Incorporating constructive explanations does not at all mean ignoring the Lorentz transformation, because
the simplest and most natural examples are, in fact, Lorentz invariant, and non-invariant
examples quickly become intractable. Exposure to constructive ideas should, rather,
increase and deepen the student's appreciation of the Lorentz symmetry, by revealing the numerous
mechanisms which are organized and brought into a coherent pattern when the symmetry holds. 

The discussion is largely qualitative, with few precise calculations,  
however those that are carried out employ natural units $c=\hbar=1$. The author claims no originality
for the calculations or demonstrations, some of which are standard; references have been cited when known. 

\section{Waves and rigidity}

Perhaps the simplest qualitative consequence of the wave paradigm, but one
which is rarely emphasized, is the non-rigidity of objects. 
Rigidity is probably the last thing one
would associate with waves, hence the simple realization that wave theories
underlie all forces (and matter as well) should help to make Lorentz contraction
a more inuitively reasonable possibility. 

Ideal rigid objects
can exist in Newtonian theory because the forces between constituents
propagate at infinite speed, allowing a perturbation in one part of the
object to be immediately reacted to by all other parts of the object. 
In reality, any change to 
the forces inside an ordinary object is propagated via electromagnetic (EM) waves,
the complete basis of non-static solutions to Maxwell's equations away from charge centers. 
But waves inherently move at finite speed, hence the forces within
an object cannot be transmitted instantaneously, and the object cannot
respond rigidly to perturbations. It will deform under an applied force, simply
because one part starts to move before the other parts experience any change
to their local EM fields. 
This fundamental non-rigidity is independent of the strength or organization of molecular bonds. 

The object necessarily deforms during acceleration, but what happens when the acceleration
stops? Does the object resume its original shape, as
happens with ordinary elastic deformations? To see that it does not requires
further discussion of the internal EM fields (Sec.\ref{sec:pot}), but it
is clear that wave propagation introduces a generic non-rigidity which is not
present in the Newtonian paradigm. 

\section{\label{sec:retard}Retardation effects in clocks}

Simple aspects of wave propagation also affect dynamic processes such as the cyclic
processes occurring in clocks. Any process built from constituent parts relies on 
wave signals traveling between the parts, and the travel time of the waves forms
part of the duration of the process. If the overall object is set into motion, the
travel time of the waves between the constituents will generically change, causing
the clock rate to change. 

\begin{figure}[h!]
\centering
\includegraphics[width=3.0 in]{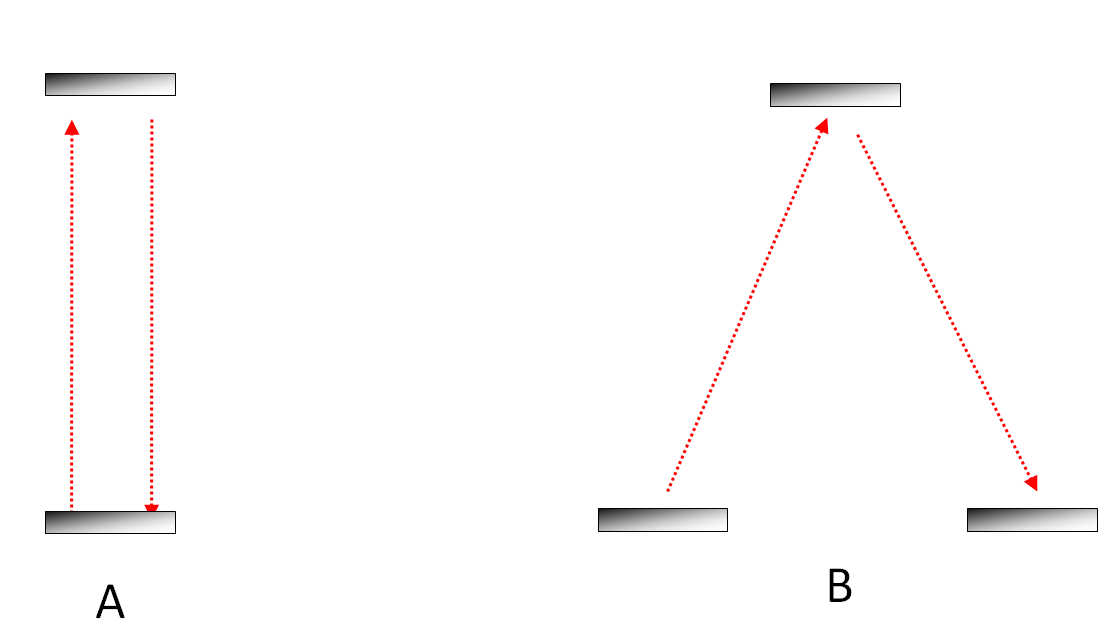}
\caption{A. Stationary light clock. B. Moving light clock (bottom mirror is drawn twice).}
\label{fig:lightclock}
\end{figure}

This effect can be seen most clearly within the venerable ``light clock'' example, which
is a model clock whose tick period is determined entirely by wave travel time. The light clock consists of
a light pulse bouncing back and forth between two mirrors (Fig.~\ref{fig:lightclock}A). 

If we now imagine the clock placed on board a spaceship with glass walls and flown by
us at high speed, it will look as in Fig.~\ref{fig:lightclock}B. The light pulse now travels
a longer distance for each cycle, hence the tick rate is slower. 

The vertically oriented clock shown in Fig.~\ref{fig:lightclock} is the most striking example, 
but other orientations also change rate, as can be seen by simple algebra. 
One doesn't get the correct relativistic answer without also considering length contraction, but 
it is nevertheless clear qualitatively that the rate must change. One may also consider
other complications, for example direction-dependence of light speed (as was originally
expected to result from motion through an aether)
and the qualitative result remains: the rate must change, except
possibly for a few particular velocities and orientations. 

Two generic aspects of wave propagation contribute to this effect. First, waves propagate at finite
speed, which both enables the construction of a light clock and also makes
retardation a factor in virtually every other kind of clock. And second, the propagation speed
of the waves does not depend on the speed of the mirrors from which they reflect; this prevents
the moving clock from automatically adjusting to restore its original rate.  
Source-independence of wave velocity should be familiar and intuitive to students, as it 
holds for all the waves studied in the basic physics curriculum. 

We note the strong contrast between source-independence, which is intuitively straightforward and
determines the qualitative fact of motion-dependent clock rates, and observer-independence, the much stronger 
assumption which is usually made. Observer-independence implies the precise Lorentz symmetry, 
but it requires a large intuitive leap because it depends on the whole range of motion-dependent phenomena occurring
within the observers themselves and their measuring devices. Students should be in a better position to
understand this postulate if they are first brought to believe that such phenomena {\it must} 
occur, as a consequence of simpler principles they have already learned.  

Source-independence is less clear-cut for dispersive waves, e.g. relativistic 
matter waves (discussed below) or the Schr\"{o}dinger waves of quantum mechanics, for which
the motion of the source can affect the frequency of emitted waves and hence their speed; 
however, this sort of complication need not deter students from developing 
basic intuitions using familiar non-dispersive waves. Even for dispersive waves
the indirect nature of the source's influence on speed makes it clear that, 
in the vast majority of cases, they will not respond to a moving source in just the 
right way to counteract retardation effects inside an arbitrarily-moving clock.

\section{Clock synchronization}

Clocks which one observer synchronizes according to a
reasonable procedure do not match those synchronized by a
differently-moving observer, using the exact same procedure. 
From the constructive point of view this phenomenon has a simple, yet subtle cause: 
synchronization of separated clocks requires additional motion to carry out. 
Two clocks must be synchronized at one location, and then one of the clocks
must be physically carried to the remote location. Carrying the clock, however, creates additional 
time dilation beyond that caused by the observer's baseline motion.\cite{edding} 

On first glance it seems that the extra effect
can be made negligible simply by making the extra motion extremely slow. However, this fails
because the slower movement takes longer to execute, hence the 
extra time dilation accumulates over a longer time. 

As with time dilation, the synchronization effect can be visualized using a
vertically-oriented light clock. We assume the clock is carried very slowly from back to front,
i.e. in the direction of the ship's velocity, covering a distance L along the
the ship (L is measured in the {\it stationary} frame). We assume that it is carried at some constant
velocity, with the initial and final accelerations having no significant effect. 

\begin{figure}[h!]
\centering
\includegraphics[width=4.0 in]{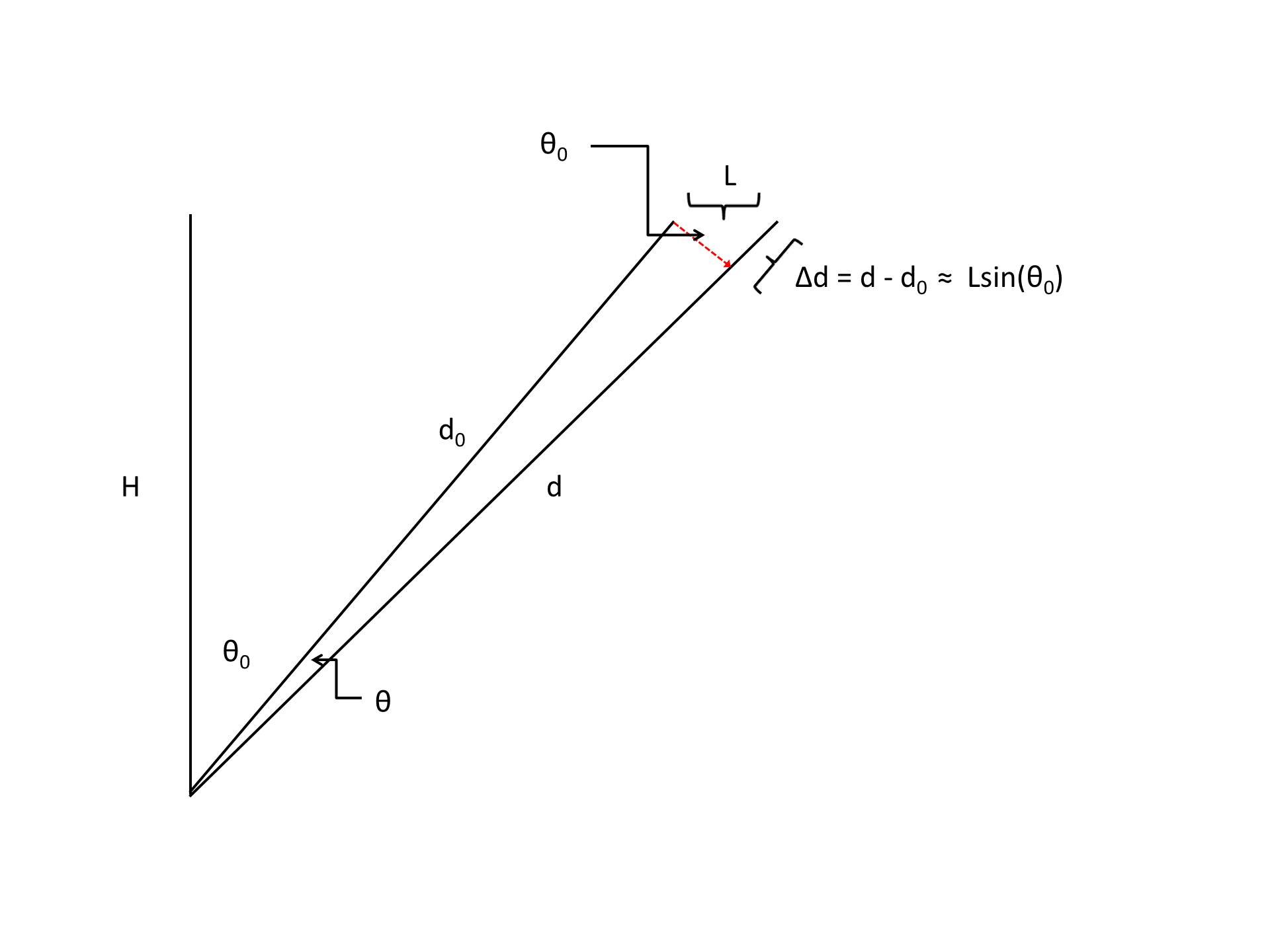}
\caption{Transported light clock.}
\label{fig:lctrans}
\end{figure}

To analyze this situation geometrically, note first that the mirror reflections have
no effect on the calculation, or equivalently one may give the light clock a height H 
such that it completes exactly one upward bounce during the time taken
to carry it the distance L. The situation is then as shown in  Fig.~\ref{fig:lctrans}.

The line tilted at angle $\theta_{0}$ shows the light pulse of a clock aboard the ship
which stays in the same place (is not carried). It completes one vertical pulse of height $H$, traversing a
distance $d_{0}$ as seen from the stationary frame. The line tilted by the
additional angle $\theta$ shows the light pulse of the carried clock, which
begins at the same location as the non-carried clock, but covers a horizontal distance
$L$, as measured in the stationary frame, and traverses total distance $d$, also as
seen the stationary frame. 

With this setup, the ``slow transport'' limit is the limit $H\to\infty$ with $\theta_{0}$ and 
$L$ held constant, and the question is whether the difference in pulse
times goes to zero, or equivalently whether the distance $\Delta d\equiv d-d_{0}$ goes to zero.
From the figure one sees that the limiting value is in fact $\Delta d = L\sin\theta_{0}$,
so it approaches zero only when the ship's speed is also zero. The carried light clock gets out
of synch with the non-carried clock, as seen by the stationary observer, even when 
the extra movement is arbitrarily slow.

The vertically oriented light clock again provides a particularly clear
demonstration because wave travel times are visualized as geometric distances. For light clocks
of other orientations, and for other forms of clock, the
synchronization effect is best demonstrated algebraically; however, the qualitative explanation 
in terms of extra motion remains the same. 

It is, of course, much more usual to consider methods of clock synchronization  
using light signals. These methods, such as the famous ``Einstein train'' thought experiment, are confusing in our view
because they conceal the fundamental physical origin of the effect. Synchronization by signalling
gets its validity from prior knowledge that light speed is measured to be the same by all observers, but
this knowledge can only be gained through experiments involving clocks and rulers which
are physically transported. Moreover, observer-independence of the speed of light
depends critically on the synchronization effect, so explanations of the latter in terms
of the former have a circular character.  We note also that, 
in field-based theories which are not Lorentz-invariant, 
synchronization by signalling becomes invalid but the physical mechanisms affecting carried clocks
are still qualitatively the same.

An essential sublety is that the synchronization discrepancy
{\it does} vanish when the observer's velocity goes to zero ($\theta_0 = 0$ on Fig.~\ref{fig:lctrans}). 
This is important since otherwise there would be no reasonable way to
synchronize clocks, even within a single reference frame. In an arbitrary field theory, 
in which Figs.~\ref{fig:lightclock},\ref{fig:lctrans} may be 
complicated by different wave speeds in different directions
and for different fields, there is, in fact, no guarantee that observers can construct useful 
reference frames by carrying clocks or any other means; the motion-dependent effects
may simply become too complex. (Of course, it is also unlikely that any 
observers would exist in such a universe.)

\section{\label{sec:packets}Wave packets, massive particles, and the relativistic mass}

The relativistic mass increase
\begin{align}
\label{relmass}
m_{rel} &= \gamma m,
\nonumber\\
\gamma &\equiv {1\over\sqrt{1-{v^2\over c^2}}}
\end{align}
seems rather ad-hoc when introduced in the context of particle mechanics.
It is difficult to understand why an elementary and indivisible piece of matter should become
harder to accelerate when moving faster. 

This effect is, however, quite natural if a particle is conceived as
a wave packet, which moreover is the way particles actually appear in modern physics.
Given the central nature of the concept and the fact that students should already be 
familiar with most of the necessary wave mathematics, we suggest that students
might benefit from seeing a wave packet description of particles during their
study of relativity. A treatment along these lines is sketched here.

Working in one spatial dimension for simplicity,
we suppose first that matter is actually
described by a wave $\xi$ obeying the simplest possible second-order wave equation,
that of transverse waves on a string:
\begin{align}
\label{m=0}
-\partial_t^2\xi + \partial_x^2\xi = 0,
\end{align}
where we have set all the constants to one for simplicity. 
All waves in this model travel with fixed speed $c=1$ regardless
of frequency, hence the model corresponds to ``massless'' waves 
such as light. To model massive particles, which is our main interest here,
one adds a linear restoring force at every point, which can be visualized
as placing Hooke's law springs on either side of the
the string at frequent intervals:
\begin{align}
\label{m>0}
-\partial_t^2\xi + \partial_x^2\xi - m^2\xi = 0.
\end{align}
Here the restoring force is labeled as $m^2$ in anticipation that $m$ will, in fact,
be understandable as the mass of a ``particle'' in this system. 

The $m > 0$ case differs qualitatively from the massless case in two important ways. First, the
presence of the ``springs'' slows down the waves, since they cannot
pass through a region of space without first exciting the springs. 
Second, the massive waves are able to sit still, at least approximately,
oscillating in place under the restoring force provided by the mass term. The massless waves, by contrast,
cannot sit still, since the only restoring force comes from neighboring parts of the string.  

The elementary wave solutions take the form
$\cos(kx \pm \omega t), \sin(kx \pm \omega t)$,
where the angular velocity $\omega$ and wavenumber $k$ satisfy
\begin{align}
\label{omegak}
\omega^2 = k^2 + m^2.
\end{align}

The elementary sine and cosine solutions don't look much like particles, since they extend
over all space, but this can be remedied by building a ``wave packet'', which is 
a superposition of waves having similar $k$ values, for example:\cite{whitham}
\begin{align}
\label{packet}
\xi(x,t) = \int_{\bar{k}-\Delta k}^{\bar{k}+\Delta k} dk \,\cos(kx - \omega t).
\end{align}
At $t=0$, the waves are all in phase at $x=0$, but interfere destructively elsewhere, creating 
a localized packet having approximate location $\bar{x}=0$ and approximate wavenumber $\bar{k}$ 
(and corresponding angular velocity $\bar{\omega}$). 
As $t$ changes, the location of the (approximate) in-phase maximum moves, and with it the wave packet. By substituting
\begin{align}
\omega \approx \bar{\omega} + {\partial\omega\over\partial k}\bigg{|}_{\bar{k}}(k-\bar{k}), 
\end{align}
one sees  that the packet moves with approximate velocity given by the ``group velocity''
\begin{align}
\label{gv}
v = {\partial\omega\over\partial k}\bigg|_{\bar{k}} = {\bar{k}\over\bar{\omega}}.
\end{align}
We henceforth drop the bar notation and use $k,\omega,x$ for the wave packet's central values. 

This ``waves on a string'' model is not so far from the real description of particles in 
modern theories. The non-particle-like extended solutions 
are also found in real quantum theories, but they are quickly converted to more localized wave packets
by interaction with other clumps of matter.\cite{mott} Quantization also prevents the waves
from simply dissipating away to zero. 

A wave packet can be subjected to a force and made to accelerate by placing it
in an external potential field, e.g. through a coupling such as $\phi(x)\xi^2$, which
is equivalent to having the ``spring tension'' $m^2$ vary with position 
(and is essentially a Higgs field coupling), 
or more complex couplings such as
$\phi(x)\xi\dot{\xi}$, which resembles the coupling of an electric potential. 

The qualitative origin of relativistic mass can be seen immediately, 
since the group velocity satisfies $v < 1$ for 
all values of $k$. The packets cannot attain this ``spring-free'' limiting speed because their propagation is
hindered by the presence of the ``springs''; hence, they must become
more and more difficult to accelerate as they approach $v=1$. In other words, their
effective mass must grow with $v$! As the limiting speed is approached, the energy applied to accelerate
the packet goes instead into internal vibrations. 

To see this in more detail, we start with the
standard expressions for energy and momentum of a vibrating string (including
the less-standard $m$ term, whose contribution is simply a Hooke's law energy):
\begin{align}
\label{EP}
\mathcal{E} &= \int dx\, {1\over 2}\left\{(\dot{\xi})^2 + (\xi^\prime)^2 + m^2\xi^2\right\}
\nonumber\\
\mathcal{P} &= \int dx\, \dot{\xi}\xi^\prime.
\end{align}
Evaluating these for a wave packet gives 
\begin{align}
\label{EPbar}
E &= \omega^2N^2
\nonumber\\
P &= \omega kN^2
\end{align}
where $N^2$ is the squared norm, $N^2 = \int dx \xi^2$. (These equations are 
actually approximate, depending on the width of the wave packet in $k-$space, but
for simplicity we will often write such equations as equalities.)

To go further in the program of constructing particles out of wave packets, 
one has to decide what value of $N^2$ constitutes a single particle. 
Not just any arbitrary convention will do, but rather it must be one
which is preserved, at a minimum, under conditions of slowly-varying external forces (``adiabatic processes''), 
because such processes should simply move particles around without creating 
or destroying them. Hence, we need to know the behavior of $N^2$ as 
the wave packet moves in a slowly-varying potential field.  

Similar questions were famously debated in the early days of quantum mechanics, and it
turns out that classical mechanics does provide a suitable adiabatic invariant,
which for the harmonic oscillator takes the form $E/\omega$.\cite{siklos}
This invariant also applies to the wave packets, because the vibrating string is essentially
a collection of harmonic oscillators, one for each $k$, as can be seen by fourier-transforming
Eq.(\ref{m>0}) in the spatial variable. 

In other words, the wavepacket norm $N$ will evolve such that
\begin{align}
\label{Nomega1}
{E\over\omega} \approx \omega N^2 \approx const 
\end{align}
and the normalization convention must be consistent with this. 
We will choose the simplest option,
\begin{align}
\label{Nomega}
N(\omega) = {1\over \sqrt{\omega}},
\end{align}
which is also the normalization arrived at through quantization.

The $1\over\sqrt{\omega}$ dependence of the normalization plays an important 
role in both relativity and quantum mechanics, and
we emphasize that it is not arbitrary, nor introduced ad-hoc in order to get relativistically
correct results, but rather is determined by the basic physics of harmonic oscillators. 
We note also that, for complex fields, in many cases it is elevated to an exact result, the
conservation of the electric charge $Q$, defined as 
\begin{align}
\label{elec}
Q = \int dx\, Im(\xi^* \dot{\xi}).
\end{align}
It is easy to verify that $Q$ is invariant, $\dot{Q}=0$, for waves $\xi$ satisfying Eq.(\ref{m>0}),
even if $m$ is taken to be a Higgs potential $\phi(x,t)$ which varies in both position and time. 

Applying this to Eq.(\ref{EPbar}), one finds for the single-particle energy and momentum
\begin{align}
\label{EPbar2}
E &= \omega
\nonumber\\
P &= k,
\end{align}
which are the well-known relations proposed by Einstein for photons, and then by
de Broglie for matter waves (in units with $\hbar = 1$). These relations are
seen to be naturally contained in the elementary physics of waves on a string.

A force $F$, by definition, acts to change the wave packet momentum by
\begin{align}
\label{F}
\dot{P} = F
\end{align}
and hence one has
\begin{align}
\label{F2}
F = \dot{k}
\end{align} 
which is exactly as expected for a force acting on the relativistic mass Eq.(\ref{relmass}),
since Eqs.(\ref{omegak},\ref{gv}) imply  
\begin{align}
\label{omegagamma}
\omega = m\gamma
\end{align}
and 
\begin{align}
k = m\gamma v.
\end{align}

Hence, the relativistic mass and its associated force law are embedded in the simple
physics of a vibrating string, which also (with many additional complications)
 is the physics of the real fields describing particles in nature. 
It is interesting also to go further and derive the standard
relativistic action for a point particle; this is done in Sec.\ref{sec:actions}.

We note for future reference that factors of $\omega$ in any expression involving
wave packets are essentially equivalent to relativistic $\gamma$ factors, as
seen in Eq.(\ref{omegagamma}).

\section{Inertially-dominated time dilation}

The relativistic mass increase, Eq.(\ref{relmass}) provides another mechanism of time dilation, since
a potential gradient which accelerates a wave packet in one direction will also reduce its
speed in orthogonal directions, decreasing the frequency of cyclical motions. This 
can be seen in a simple ``clock'' consisting of a circular orbiting system, accelerated
slowly along the axis perpendicular to the orbit. 

The tangential force applied to the orbiting particle is zero, hence the orbital momentum doesn't
change, but since the effective mass does increase one finds that the orbital speed is reduced:
\begin{align}
\label{tangential}
0 &= {d\over dt}(m \gamma v_{\perp}) = m{d\over dt}(\gamma v_{\perp})
\end{align}
implying $\gamma v_{\perp} = const$, or
\begin{align}
\label{tangential3}
v_{\perp} \propto {1\over\gamma}.
\end{align}
Assuming that the orbit radius does not change (as also implied by conservation of angular
momentum), this implies that the orbital period increases by the time dilation factor $\gamma$. 

Actually we have been slightly imprecise, since the particle's $\gamma$-factor 
differs from that of the overall system. Denoting the system's overall velocity as 
$\vec{\beta}$, one has $\vec{\beta} \perp \vec{v_\perp}$, and
\begin{align}
\bar{\gamma} &= (1 - (\vec{\beta} + \vec{v}_\perp)^2)^{-{1\over 2}}
\nonumber\\
&= (1 - \beta^2 - v_\perp^2)^{-{1\over 2}}
\nonumber\\
&= \gamma (1 - \gamma^2 v_\perp^2)^{-{1\over 2}}.
\end{align}
However, this leads to the same conclusion as before, since the force equation 
$0 = {d\over dt}(m \bar{\gamma} v_{\perp})$
still depends on $\gamma$ and $v_\perp$ only through the combination $\gamma v_{\perp}$. 

Similar reasoning applies to other clocks using inertial mechanisms, for example 
a vertically-oriented ``bouncing ball'' clock, the massive analog of the light
clock. As the ship holding the clock slowly accelerates transverse to the bounce directions, 
the same acceleration must be applied to the ball to keep it bouncing between the two plates. The clock's
caretakers on board the ship must do this without applying any vertical force,
because that invalidates the system's timekeeping function even within their own reference frame. 
Hence they will maintain the clock through small impulses perpendicular to its bounce 
direction, leading to the same slowing effect seen with the accelerated orbit. 

For general orientations (non-transverse), the calculations are much more complex and
involve not just the relativistic mass, but also changes to the potential fields, as
described in the following section. One must certainly resort to Lorentz invariance to 
prove precise time dilation for an arbitrary system (and conversely, the precise time-dilation formula
will not be true for systems lacking Lorentz invariance). However, the qualitative mechanisms can be understood
from the simpler configurations.

\section{\label{sec:pot}Changes to potential fields}

We have not yet addressed length contraction directly, and it is indeed difficult 
to demonstrate constructively, at least in a quantitatively correct fashion. 
The problem is that lengths of objects are determined by the shapes of quantum
orbitals, whose calculation about a moving center of force is very difficult without
resorting to Lorentz invariance. Even in the classical limit of particle orbits 
the only tractable example appears to be the transversely-oriented circular orbit discussed above. 

The qualitative origin of length contraction is, however, very easy to see, and has been recognized
since Fitzgerald's famous ``letter to Science'' in 1889.\cite{brown2} It is caused by
the change in shape of potential fields about a moving source, a change which is made
inevitable by basic aspects of the wave propagation by which the field is generated. Rather
than rigorously derive the shape change, which is done in standard textbooks,\cite{jackson} or
attempt to explicitly construct length contraction from it,
we will be content to describe how the moving-source potential connects to familar aspects of wave
propagation. It is hoped that this will make it more intuitively obvious that the shape of the potential
field must change, from which it is fairly inescapable that the shape of any object built using
such fields will also change. 

We consider first a scalar field, which is essentially a Higgs field, and 
produces a standard $1/r$ potential in three dimensions. 
The field generated by a source moving with constant velocity $\vec{\beta}$ 
has the following form:
\begin{align}
\label{movingphi}
\phi = {1\over \gamma R (1-\vec{\beta}\cdot\hat{R})}.
\end{align}
The three factors in the denominator come about as follows. The $\gamma$ factor reflects the decrease
in amplitude of the source wave packet as it is accelerated to higher speed, Eq.(\ref{Nomega}) (note 
that $\phi$ will typically be sourced by the wavepacket norm $N^2$ through a coupling such
as $\phi(x)\xi^2(x)$). 
The $R$ factor is the familiar $1/r$ falloff for a potential field in three dimensions, 
but measured from the retarded emission point along the trajectory of the source; this retardation correction 
ultimately reflects the fact that the potential is generated by emission of spherical 
waves from the source, and waves inherently move
at finite speed. The final factor, $1-\vec{\beta}\cdot\hat{R}$, can be thought of as
an amplitude component of the Doppler effect,
which persists for zero-frequency fields (here $\hat{R}$
is a unit vector in the retarded direction, along which $R$ is measured).
This factor can be seen most clearly in the simplest of all wave equations, 
the first-order equation in one dimension
$\partial_t\phi + \partial_x\phi = 0$, for which the solution for a right-moving source is 
$\phi = {\theta(x - \beta t) \over{1 - \beta}}$.

Taking motion in the $\hat{x}$ direction, $\vec{\beta} = \beta\hat{x}$,
one can work out the geometry of retarded distance and verify that the
potential is just the Lorentz-transformed $1/r$ potential,
\begin{align}
\label{ltransphi}
\phi(\vec{x},t) = {1\over\sqrt{\gamma^2(x-\beta t)^2 + y^2 + z^2}}.
\end{align}
Hence one sees that this potential arises through generic wave mechanisms, in particular finite
speed and the Doppler effect. 

The EM fields about a moving source can be derived from the scalar field solution. The 
EM scalar $\phi_E$ has nearly the same form as the Higgs (which we will call $\phi_H$). 
The only difference is that the EM scalar couples to the adiabatically conserved quantity $\omega N^2$ 
(which for charged particles becomes the exactly-conserved electric charge Eq.\ref{elec}), 
hence is not diminished by the extra $1\over\gamma$ factor seen in Eq.(\ref{movingphi}). 
We can write then
\begin{align}
\label{twophis}
\phi_E = \gamma\phi_H.
\end{align}

The EM vector potential $\vec{A}$ is generated in exactly the same way as $\phi_E$, except that it
couples to the velocity of the source. Hence for a moving point source one has
\begin{align}
\label{A}
\vec{A} = \vec{\beta}\phi_E = \vec{\beta}\gamma\phi_H 
\end{align}
resulting in electric field
\begin{align}
\label{E}
\vec{E} &= -\vec{\nabla}\phi_E - \partial_t \vec{A}
\nonumber\\
&= -\vec{\nabla}\phi_E +\vec{\beta} (\vec{\beta}\cdot\vec{\nabla})\phi_E.
\nonumber\\
&= -\gamma\vec{\nabla}\phi_H + \gamma \vec{\beta} (\vec{\beta}\cdot\vec{\nabla})\phi_H. 
\end{align}

The gradient of the Higgs field, $\vec{\nabla}\phi_H$, does not remain radial about the
moving center, but the electric field given by Eq.(\ref{E})
is radial, as can be seen by substituting Eq.(\ref{ltransphi}). 
The shape of the moving electric field, first calculated by Heaviside in 1888,
is shown in Fig.\ref{fig:movingcharge} below.

\begin{figure}[h!]
\centering
\includegraphics[width=4.0 in]{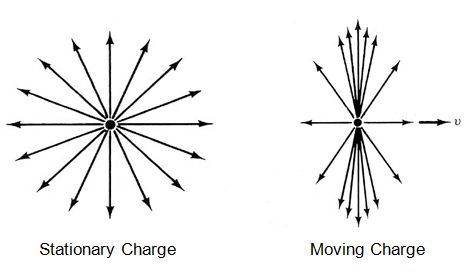}
\caption{Electric field about a uniformly-moving source.}
\label{fig:movingcharge}
\end{figure}

This famous and striking image shows the moving electric field ``compressed'' along the direction
of motion, which certainly suggests that orbits and orbitals within the potential should 
experience similar contraction; at a minimum, they should undergo {\it some} shape change. 
It is not hard to see the contraction approximately, 
for slowly-orbiting particles (Larmor and Lorentz did this during the decade prior to Einstein's work), 
but it is even easier to verify full Lorentz invariance and apply that. We do not attempt a more detailed
treatment of orbits (or quantum orbitals) about a moving center, but we nevertheless 
feel confident in asserting, with Harvey Brown, that ``shape deformation produced by motion is far from the proverbial
riddle wrapped in a mystery inside an enigma''.\cite{brown}

A precise construction of length contraction may not be essential for understanding the phenomenon, but
students can still benefit from attempting an orbit calculation, if only to get a sense of just how complicated
even this simple problem becomes in the wave-based setting. The first-order calculation is a simple exercise in 
perturbation theory, and one can also simulate the accelerating system on a computer, as suggested by Bell in his
original manifesto.\cite{bell} A creative scenario due to Miller also allows the contraction to be
seen without massive complication.\cite{miller}

\section{\label{sec:decays}Time dilation of decays}

Although usually treated within quantum field theory, the relativistic aspects of particle decays can
be largely understood semiclassically. Indeed the essential factor can be seen already in the 
driven harmonic oscillator, 
\begin{align}
\label{dsho}
\ddot{x} + \omega^2 x = F. 
\end{align}
Taking $F = \theta(t)\cos{\omega t}$ gives the solution
$x = {t\over2\omega}\theta(t)\sin{\omega t}$,
showing that the amplitude increase is damped by a factor $1\over\omega$. The
same factor is seen more generally in the Green function 
\begin{align}
\label{shogreen}
G(t-t^\prime) = {1\over\omega}\theta(t-t^\prime)\sin{\omega(t-t^\prime)}. 
\end{align}

The $1\over\omega$ suppression carries over directly to particle decays since, as noted above, the particle
fields are just collections of harmonic oscillators having $\vec{k}$-dependent frequency. 
Decays occur when one field drives one or more other fields at their resonant frequencies, 
and the factor of $1\over\omega$ is essentially
the relativistic dilation factor for moving wavepackets, as noted at the end of
Sec.\ref{sec:packets}. (By ``resonant frequencies'' in this
context, we mean that both $k$ and $\omega$ should satisfy Eq.\ref{omegak} for the field being driven.)

The simplest example comes not from true decay but rather oscillation between 
two particles having the same mass. In terms of
the one-dimensional string model this corresponds to two parallel, identical strings 
attached to each other by additional springs running between them. 

Letting $X,Y$ be the displacements
of the two strings, the springs running between them create a Hooke's law interaction energy
${1\over 2}g(X-Y)^2$. Expanding this, the $X^2$ and $Y^2$ terms just shift the 
mass $m^2$ of each string (recalling Eq.\ref{m>0}), leaving the effective interaction term $gXY$. 
The coupled equations of motion then take the form
\begin{align}
\label{twostring}
\ddot{X} - X^{\prime\prime} + m^2 X = gY
\nonumber\\
\ddot{Y} - Y^{\prime\prime} + m^2 Y = gX.
\end{align}

A complete basis of solutions is easily found by taking $X$,$Y$ exactly in or out of phase, $X = \pm Y$. 
The out-of-phase modes stretch the springs connecting the two strings, hence
have higher oscillation frequencies than the in-phase modes, and one finds  
\begin{align}
\label{Omega}
\Omega_{\pm} &= \sqrt{k^2 + m^2 \mp g}
\nonumber\\
&\approx \omega \mp {g\over 2\omega},
\end{align}
where the last line is the small-$g$ approximation.

To build a wave packet which initially exists only on the $X$ string, one can combine
two identical packets built using the in-phase and out-of-phase modes. The packets initially
cancel each other on the $Y$ string, but because they have slightly
different frequencies $\Omega_\pm$, the initial ``particle'' 
will oscillate between the two strings at the beat frequency 
\begin{align}
\label{beat}
\omega_{osc} &= {1\over2}(\Omega_- - \Omega_+) 
\nonumber\\
&= {g\over 2\omega}
\nonumber\\
&= {g\over 2m\gamma}.
\end{align}
The last line shows the relativistic time dilation of the particle oscillation frequency (and used
Eq.\ref{omegagamma}). We note that the two packets will also separate over time, due to their
differing group velocities, however this effect is of order $O({1\over{\omega^2}})$.

True multiparticle decays are more involved, but a semiclassical treatment 
can still be given showing the relativistic factors; this is done in Sec.\ref{sec:multi}.

\section{Dynamical evolution of states}

One aspect of relativity which is not always clear in the axiomatic viewpoint is which objects 
actually attain their Lorentz-transformed form after a dynamical acceleration process.  
Assume a system begins at rest in a state $S$ and is adiabatically accelerated to velocity $v$, 
resulting in a state $S^\prime$. Is this state identical with the state $L(S)$ obtained by
Lorentz-transforming $S$ to velocity $v$?

To make any statements about $S^\prime$, its evolution must be constrained by some
quantity which is at least adiabatically invariant. To connect this to a Lorentz transformation, the
same quantity must be Lorentz invariant. Such quantities can be found by looking at the 
phase and nodes of a standing wave; indeed, by counting cycles or nodes along
well-defined lines or closed curves, one obtains integers which should be adiabatically invariant. Moreover, they 
will also be Lorentz invariant, since any proper Lorentz transformation can be built up
``adiabatically'' by applying numerous small transformations in succession. If these suffice to
define the state $S$, then one must have $S^\prime = L(S)$. 
Bound states in particular are constrained this way, since they are
essentially three-dimensional standing waves, and since bound state solutions define the shape of objects, 
it follows that most objects will dynamically contract. 

An example of distances which don't contract are the ``disconnected'' distances discussed by Miller,
e.g. the distance between two separate spaceships.\cite{miller} If the ships begin at rest, with original distance $D$, 
and then accelerate to speed $v$, the final distance $D^\prime$  will not, in general, be equal to
the contracted value $D\gamma_v^{-1}$; rather, it depends on the details of the acceleration 
(e.g., the rear ship could accelerate first and crash into the one in front!) If, however, the
{\it observer} accelerates, then $D^\prime$ will contract as expected, because the real effect is
occurring within the observer's own measuring devices, and hence applies uniformly to any distance which that
observer subsequently measures (cf. Sec.\ref{sec:princ}).

Another example of objects which don't dynamically contract are the wavepackets 
themselves. To see this, consider the one-dimensional massive wave equation with a weak linear potential 
\begin{align}
\label{weqwithpot}
\ddot{\xi} - \xi^{\prime\prime} + m^2 \xi = -fx\xi
\end{align}
and fourier-transform in the spatial variable to obtain
\begin{align}
\label{weqft}
\partial_t^2\hat{\xi} + k^2\hat{\xi} + m^2 \hat{\xi} = -if\partial_k\hat{\xi}.
\end{align}
For small $f$ one has approximately the free-field behavior
$\partial_t\hat{\xi}(k,t) \approx -i\omega_k\hat{\xi}(k,t)$, hence we can write
a  wave packet solution in the form
$\hat{\xi}(k,t) = \psi(k,t)e^{i\omega_k t}$, where the envelope $\psi$ is slowly-varying in time
compared to $e^{i\omega_k t}$.  Substituting into Eq.(\ref{weqft}), and approximating 
$|\partial^2_t\psi| \ll |\omega_k\partial_t\psi|$, gives
\begin{align}
\label{weqft2}
0 &= 2i\omega_k\partial_t\psi - \omega_k^2\psi + k^2\psi + m^2 \psi +if\partial_k\psi 
			-ft\psi{\partial\omega_k\over\partial k}
\nonumber\\
&= 2i\omega_k(\partial_t + {f\over2\omega_k}\partial_k +{ifv_k \over 2\omega_k}t )\psi
\end{align}
where $v_k \equiv \partial_k\omega_k = {k\over\omega_k}$. 
This is solved by a $k$-space envelope which moves with time, plus a phase factor 
(the latter can be calculated as a power series in $t$; the lowest order is shown):
\begin{align}
\label{wpktspread}
\psi = \chi(k - {f\over2\omega_k}t)e^{-i{fv_k\over4\omega_k}t^2}.
\end{align}

As time progresses, the wave packet momentum therefore evolves approximately by 
\begin{align}
\label{kevolve}
{\dot{k} = {f\over2\omega_k}}, 
\end{align}
where we focus on the case $f > 0$, $k > 0$, i.e. where the force is acting
to increase the momentum.
Since $\omega_k$ increases with $k$, this means that the packet is narrowing in $k$-space, hence
broadening as a function of $x$; the packet expands rather than contracting (this is in addition
to the natural spreading which every wave packet experiences over time). 

The behavior changes if the potential is coupled to time derivatives of the field, so that
$f$ depends on $\omega$. If $f \propto \omega$, the packet neither expands nor contracts moving in 
the potential field, but if $f = 2\omega^2$ (choosing the proportionality constant for convenience), 
the packet does contract relativistically, as can
be seen by analyzing the change in time of its width in $k$-space (starting from Eq.\ref{kevolve}, applied
to the momentum spread $\Delta k$ of the packet):
\begin{align}
\label{pktcont}
\dot{k} &= \omega_k
\nonumber\\
\dot{\Delta k} &= \Delta\omega_k = {\partial\omega\over\partial k} \Delta k = v\Delta k
\nonumber\\
\end{align}
where $v$ is the packet group velocity. Also (again using Eq.\ref{kevolve})
\begin{align}
\label{align}
\dot{\omega} = {\partial\omega\over\partial k}\dot{k} = \omega v
\end{align}
which is simply the Newtonian expression $\dot{E} = Fv$ for energy change in a force field. 
Combining the equations gives
\begin{align}
\label{pktcont2}
{\dot{\Delta k}\over\Delta k} = {\dot{\omega}\over\omega}
\end{align} 
which integrates to yield
\begin{align}
\label{pktcont3}
{\Delta k} \propto \omega = m\gamma
\end{align} 
showing that the packet's width in $k$-space expands by the relativistic factor, implying
relativistic contraction of the width in position space. 

Since $f \propto \omega^2$ corresponds to a force which couples to energy (recalling 
the $\dot{\xi}^2$ term in the expression for the energy of the vibrating string,
 Eq.\ref{EP}), i.e., to a gravitational force, the contraction
might have been expected in this case, in view of the equivalence principle.

\section{\label{sec:emc}Mass and energy}

In a world described by field theories, everything is ``energy'', hence understanding the
mass/energy relationship mainly means understanding where mass comes from in the field theory context. 
According to Einstein, mass should be just another word for the energy content of the field
configuration, so the question becomes how this actually translates into inertia, i.e., 
resistance to acceleration.  

One of the primary mechanisms was already seen in Sec.\ref{sec:packets}, namely the intrinsic
mass of a field. This can be visualized as arising from ``springs'' having force constant $m^2$, 
which both impede the motion of wave packets, i.e. increase their inertia,
and increase their internal frequency of oscillation, i.e. their energy. This is exactly in accordance
with Einstein's formula (recall Eq.\ref{EPbar2}; for $k=0$ one has just $\omega=m$). 

Closely related to this is the relativistic mass effect, by which kinetic energy of motion 
results in an increase of effective mass. This was also seen in the waves-on-a-string model
(cf. Eqs.\ref{EPbar2},\ref{omegagamma}), and is implicit in the fact that wave packets
in the presence of ``springs'' (i.e, having $m > 0$) 
can never exceed the speed of spring-free packets ($m = 0$). Hence the packets must become harder to accelerate with
speed, i.e., their inertia must grow. The energy which one applies to a wave packet
ends up going, not into speed, but rather into internal oscillations of the field; the particle resembles
an old, rattletrap car which vibrates badly at high speeds and can't be further accelerated. 

The other principle mode of mass generation in field theories was actually the first piece of the relativity
puzzle to be glimpsed, by J.J. Thomson, in the year of Einstein's second birthday.\cite{thomp} Thomson
realized that the EM field of a charged particle will act back on the particle itself,
resisting its acceleration and adding to its inertia; part of this effect is just the familiar phenomenon of
self-induction. This ``back-reaction'' mechanism, operating in the strong nuclear force, in fact 
is the source of nearly all inertia in the everyday world, only a small fraction
of which arises from intrinsic masses (or Higgs mechanism).\cite{wil1}

Back-reaction, plus relativistic mass, account for most examples of mass/energy connection.
To take a concrete example, the energy of an excited state of hydrogen is greater than that of the
ground state, hence the excited state mass must also be greater. 
The additional energy is EM potential energy, but how
can that physically act as inertia, i.e., how can it impede the acceleration
of the atom? The only possible mechanism is through backreaction 
of the internal EM field onto the charged particles. The EM portion
of the mass difference is then accounted for by differing backreaction
from the differing internal fields of the two states.\cite{masspaper} 
(We note also that the additional EM energy in the excited state is partially offset
by a decrease in the electron kinetic energy, which translates
into a lower contribution to the atom's overall inertia, by the relativistic mass effect.)

Eventually the excited state decays back to the ground state, emitting a photon and lowering its mass.
Since the extra mass of the excited state was due originally to EM potential energy,
 this ``conversion'' of mass to energy is actually nothing but conversion of
electromagnetic field energy to a different form.

\section{\label{sec:maxspeed}Maximum speed}

One of the most common questions asked about relativity is ``why can't anything move faster than light?'' 
To such a simple question one would expect a simple answer, but
the most common answer is anything but. Typically one is told that superluminal signal
propagation would allow backwards time travel, resulting in paradoxes, hence impossibility. 
In our view this answer, although not exactly incorrect, is not very informative
since it provides little insight into the physical mechanisms at work.  

Another standard answer, providing somewhat greater insight, is to note that the 
relativistic mass of an object diverges as it approaches light speed, making it physically impossible to exceed
this speed. However, this answer only applies to objects with mass, and it
also begs further explanation of the relativistic mass effect itself. 

In our view, a more comprehensible and arguably more accurate answer is to appeal directly to the underlying
wave nature of the universe, and to the questioner's intuitive understanding of wave
phenomena. From everyday experience with water waves, people understand that 
waves cannot be made to move faster than their natural speed. One can't speed up an ocean wave by
pushing on it; such an effort merely creates more waves, not faster waves. The sound from
a jet airplane or rifle shot travels the same speed as that from a turtle. The reason
nothing in our universe moves faster than light is then simply that everything in the universe is 
built out of waves that have that speed. The massless waves
move at exactly $c$, exactly analogous to familiar waves, while the massive ones
are hindered by the added ``springs'' (recalling Sec.\ref{sec:packets}) and hence move at lower speeds.  

Implicit in this explanation, of course, is the idea that {\it everything} is made of waves, so that
there is no form of motion in the universe other than wave motion. There is no analog 
of ``supersonic'' travel or Mach numbers for the fundamental waves making up the universe, 
because these refer to the speed of objects {\it other} than sound waves; an object
which was somehow made from sound waves would be inherently unable to travel faster 
than the speed of sound, and likewise the objects in our universe cannot travel faster
than the waves from which they are made. \cite{shupe}

The question ``why can't anything move faster than light'' is then seen to be equivalent
to asking why sound waves don't move faster than the speed of sound. This answer,
although tautological at first glance, in fact reflects the true, underlying origin of
the cosmic speed limit. 
It is deeper than the second answer, because the relativistic mass effect itself 
comes from underlying waves having bounded group velocity, as was described in 
Sec.\ref{sec:packets}. It is also more informative
than the first answer, because it places the phenomenon of bounded speeds within the 
wider physical context of wave motion, rather than the very specific context of Lorentz invariance,
which entails specific combinations of fields all having the same speed constant.

\section{\label{sec:princ}Connecting to the principle approach}

If everything is ``made out of waves'', then motion-dependent changes to objects and
processes are generic and inevitable, as we have tried to illustrate in the previous sections. 
Students should not have trouble understanding these generic
effects, since they connect to the most elementary types of waves studied in every curriculum. 

Students should then be ready to take the next step, realizing that measurement itself makes use of various objects
and processes, hence measurements made by differently-moving observers must also differ. 
This conclusion sounds quite dramatic when derived abstractly from axioms, but 
the pedagogical goal should be for students to think of it as just an obvious consequence of basic
physical mechanisms that they already understand. 

Realizing that all objects, including measuring devices, must undergo motion-dependent changes, the
next logical question is whether some order can be brought to what seems like a potentially very
complex or chaotic situation. How is it that we humans can consistently measure things, or even exist at
all, given that motion affects all things, and that our planet is certainly moving? At this point students
should be ready to more clearly understand the content and importance of the Lorentz symmetry
and the principles it embodies. These may, indeed, make a greater impression on students when
introduced through this longer route, as they will
already have an idea of the vast range of complexities which can arise from 
wave-based underlying theories, and will be better able to appreciate both how surprising and how essential
the Lorentz symmetry really is.

\section{\label{sec:actions}Appendix: relativistic particle action and wave phase}

To cement the connection between the particle-oriented approach to relativity and 
the wave packet derivations provided in Sec.\ref{sec:packets}, it is worthwhile to show how the customary
relativistic particle action\cite{gold} 
\begin{align}
L = -\int {m\over\gamma} 
\end{align}
arises from the fundamental wave theory. 
For simplicity we take the wave field $\xi$ to be complex, which permits exponential solutions
rather than sines and cosines. In terms of the string picture this just means
allowing the string to vibrate in both transverse directions, although
the ``springs'' creating the mass term become hard to visualize.

The standard action for a vibrating string, with the added Hooke's law mass term, is 
\begin{align}
\label{packetact}
S = \int dtdx\, \{ |\dot{\xi}|^2 - |\xi^\prime|^2 - m^2|\xi|^2\}.
\end{align}
The Euler-Lagrange equation of motion following from this action is just Eq.(\ref{m>0}).

With constant $m$ there is no applied force, which results
in trivial constant-speed motion for the wave packets. As mentioned in Sec.\ref{sec:packets}, one can include a
force from an external Higgs-like potential simply by allowing $m$ to vary with $x$ and/or $t$.  
We will continue to work in one dimension, although the results carry easily
to three dimensions by converting $k$ to a vector $\vec{k}$. 

We seek an action which will govern wave packets for which $k$ and $\omega$ vary
slowly in time, i.e. for which $\dot{k},\dot{\omega}$ are treated as small
parameters and retained to first order. 
The wave packet can be written as a plane wave times a 
slowly-varying (complex) envelope function, leading to 
the ansatz  
\begin{align} 
\label{wpansatz}
\xi &= \exp i(kx - \omega t)F(x,t)
\nonumber\\
\dot{\xi} &= [(-i\omega -it\dot{\omega} + ix\dot{k})F + \partial_tF]\exp i(kx - \omega t), 
\nonumber\\
\xi^\prime &= (ikF + \partial_xF)\exp i(kx - \omega t), 
\end{align}
and the action becomes (discarding terms quadratic in derivatives)
\begin{align}
\label{act2}
S \approx \int dtdx \{(\omega^2 + 2t\omega\dot{\omega} - 2x\omega\dot{k} - k^2 - m^2)|F|^2
				-2 Im(F^*(\omega\partial_tF + k\partial_xF))\}
\end{align}

To reduce this to an action governing the ``slow'' changes $\dot{k},\dot{\omega}$, we  
apply as constraints the ``fast'' consequences of the wave equation,
$\omega^2 = k^2 + m^2$ and $N^2 = {1\over 2\omega}$, where the latter is Eq.(\ref{Nomega}) 
with an extra factor appropriate to the complex field.\cite{quantnorm}
An additional constraint is $F(x,t)=F(x - vt)$, expressing 
the fact that a wave packet built from modes satisfying the first constraint
will move at the group velocity $v={k\over\omega}$; this eliminates the
term involving derivatives of $F$. Then inserting the first constraint gives
\begin{align}
\label{act3}
S &\approx \int dtdx (2t\omega\dot{\omega} - 2x\omega\dot{k})|\xi|^2.
\end{align}

Assuming the packet is relatively narrow, we can approximate
$x$ by the packet center location and perform the $dx$ integral, 
leading to 
\begin{align}
\label{act4}
S &\approx \int dt (2t\omega\dot{\omega} - 2x\omega\dot{k})N^2
\nonumber\\
&= \int dt(t\dot{\omega} - x\dot{k})
\nonumber\\
&= \int dt (-\omega + \dot{x}k)
\nonumber\\
&= -\int dt {m\over\gamma},
\end{align}
which is the customary relativistic particle action 
(the last step used the group velocity $\dot{x} = {k\over\omega}$; the third step 
integrated by parts, discarding the boundary terms, 
which don't affect the equations of motion arising from the action).  

The derivation highlights the connections between the phase of the
underlying wave, the particle's proper time, and Lorentz invariance.
The third line can be written in terms of the phase $\theta$ as
\begin{align}
\label{phase}
S &= \int dt (-\omega + \dot{x}k) = \int dt ({\partial\theta\over\partial t} + \dot{x}{\partial\theta\over\partial x})
\nonumber\\
&= \int dt {d\theta\over dt} = \int d\theta.
\end{align}
Hence the variational principle governing the particle's motion is to minimize (or extremize) the 
number of cycles of its underlying wave. 
Also, Lorentz invariance of $S$ is equivalent to invariance of the phase, which
can be seen in customary 4-vector form by noting
\begin{align}
\label{4vec}
d\theta = -\omega dt + \vec{k}\cdot d\vec{x} = k^\mu x_\mu.
\end{align}
Lastly, from the last two steps of Eq.(\ref{act4}) the particle's proper time and
phase are related by
\begin{align}
\label{proptime}
m d\tau = d\theta. 
\end{align}

\section{\label{sec:multi}Appendix: Multiparticle decays}

Multiparticle decays require quantum mechanics for a correct treatment, 
but a heuristic sketch shows the important relativistic factors. We extend 
the previous two-field case of Sec.\ref{sec:decays} 
to let a field $X$ of mass $m$ interact with $N$ others
$X_1,...,X_N$, having masses $m_i$, with interaction $\mathcal{L}_I = gXX_1X_2\cdot\cdot\cdot X_N$. 
We consider an initial wave packet in $X$, which we will assume
to be narrow enough in momentum space to be treated as a 
pure mode with fourier components $k,\omega$.

The packet on $X$ will also have amplitude proportional to $1\over\sqrt{\omega}$,
in accordance with Eq.(\ref{Nomega}); indeed the main concern will be
to show how the various $\omega$ and $\omega_i$ factors combine to create
a time-dilated decay rate. Details not connected to this, e.g. whether
the fields are real or complex, will be ignored.

Classically, a solution with $X_i = 0$ for all $i$ is stable and no
decay occurs. In quantum mechanics, however, a field is never truly zero, but contains irreducible 
zero-point fluctuations, which we will label $\bar{X_i}$. No energy can be drawn from
these fluctuations, but they can combine with true excitations to create resonances which do transfer energy. 
Focusing on $X_1$, for example, its equation of motion is
\begin{align}
\label{X1}
\ddot{X_1} - X_1^{\prime\prime} + m_1^2 = gXX_2\cdot\cdot\cdot X_N
\end{align}
and one sees that the zero-point fluctuations $\bar{X_2},...,\bar{X_N}$ can modulate the $X$ wavepacket to
create a resonant driving term for $X_1$, transferring energy into that field. 

The ``vacuum'' of zero-point fluctuations must be stable under adiabatic changes in external potentials, hence
the amplitudes of these fluctuations must be governed by Eq.(\ref{Nomega}), so each mode of frequency $\omega_i$ 
comes with an amplitude factor $1\over\sqrt{\omega_i}$, and we can write schematically
\begin{align}
\label{barX}
\bar{X_i} = \int {dk_i \over \sqrt{\omega_i}} e^{i(k_ix - \omega_i t)}.
\end{align}

As in the driven harmonic oscillator, an exact resonance with an $X_1$ mode $k_1,\omega_1$ leads to
a linearly increasing amplitude, damped by $1\over\omega_1$ (recall Eqns. \ref{dsho},\ref{shogreen}). 
Hence, assuming that the $X$
wavepacket comes into existence at $t=0$ (or the interaction ``turns on'' at that time), 
as a first approximation one expects that each $X_1$ mode will grow as
\begin{align}
\label{onemode}
X_1^{k_1,\omega_1} \propto {gt\over\omega_1\sqrt{\omega}}\int {dk_2\cdot\cdot\cdot dk_N \over \sqrt{\omega_2\cdot\cdot\cdot \omega_N}}
					\delta(k - k_1 - k_2 - \cdot\cdot\cdot - k_n)	\delta (\omega - \omega_1 - \omega_2 - \cdot\cdot\cdot - \omega_n).
\end{align} 
Here the delta functions express the conservation of energy and momentum. 

Next we note that the time dilation of the decay rate of $X$ takes a simpler form when expressed in terms
of energy. The single-particle $X$ wavepacket has energy $\omega = m\gamma$, 
so a decay rate proportional to $\gamma^{-1}$ is equivalent
to a rate of energy loss which is independent of $\omega$. To put this another way, if the rate of
energy loss is independent of $\omega$, then the time taken to lose half of the starting
energy is proportional to $\omega$, i.e., to $\gamma$.
Accordingly we will study the rate of energy emission into the fields $X_i$, rather than looking
at the amplitudes directly, as would be done in a quantum mechanical treatment. 

Looking again at $X_1$, by Eq(\ref{EPbar}) its energy is given in terms of the mode amplitudes as
\begin{align}
\label{X1E}
E_1 = \int dk_1\, \omega_1^2 \left|X_1^{k_1,\omega_1}\right|^2.
\end{align}
In view of Eq.(\ref{onemode}) it is tempting to say the energy should grow as $t^2$, and this is correct
for a problem with exactly matching modes, such as the $X,Y$ case 
of Sec.\ref{sec:decays}; indeed $t^2$ is the first term in the expansion of $\sin^2(t)$ 
for the energy transferred to the $Y$ string in the oscillating solution. 

Multiparticle decays don't have exactly matching modes, but rather
each $X$ mode is smeared by interaction with the $\bar{X_i}$, 
and the $X_1$ equation of motion then amplifies that vanishingly small subset of modes 
satisfying $\omega_1^2 = k_1^2 + m_1^2$. This would imply a rate of zero, were it not
for an additional subtlety (here we are giving a heuristic explanation of ``Fermi's Golden Rule''
of quantum mechanics). Because the interaction begins at a particular time $t=0$, 
the frequency $\omega_1$ is not well-defined at the outset, but rather is subject
to the ``energy-time'' uncertainty principle 
\begin{align}
\label{ET}
\Delta\omega_1 \propto {1\over t},
\end{align}
which arises because it takes time for destructive interference to weed
out the non-resonant modes (as seen explicitly writing the solution in terms of the Green function). 
The response is therefore increased at early times due to contributions from
a wider range of driving modes, but for late times the contributions of modes not satisfying 
$\omega_1^2 = k_1^2 + m_1^2$ increasingly cancel out, so the response comes from an ever-narrower subset of
the whole $k_2,k_3,...,k_N$ space. The net energy emitted into $X_i$ is then proportional to $t$, not $t^2$ (or zero). 

The full result result for $E_1$ then looks as shown below, where again the main goal is to 
track the various $\omega$ factors which create the relativistic
dilation effect. We note that
the factors of $\omega_1$ cancel, and that each factor of $1\over\sqrt{\omega_i}$ for $i > 1$
is squared, as is the factor of $1\over\sqrt{\omega}$:
\begin{align}
\label{X1E2}
E_1 \propto {{g^2 t}\over\omega}\int dk_1 
				\int {dk_2\cdot\cdot\cdot dk_N \over {\omega_2\cdot\cdot\cdot \omega_N}}
					\delta(k - k_1 - k_2 - \cdot\cdot\cdot - k_N)	\delta (\omega - \omega_1 - \omega_2 - \cdot\cdot\cdot - \omega_N).
\end{align}
Writing $dk_1 = \omega_1{dk_1\over\omega_1}$ and adding together all of the $E_i$ gives the total energy emission 
\begin{align}
\label{allE}
\Delta E &\propto {g^2 t}
				\int {dk_1\cdot\cdot\cdot dk_N \over {\omega_1\cdot\cdot\cdot \omega_N}}
				\left({{\omega_1 + \cdot\cdot\cdot + \omega_N}\over \omega}\right)
					\delta(k - k_1 - k_2 + \cdot\cdot\cdot - k_N)	\delta (\omega - \omega_1 - \omega_2 - \cdot\cdot\cdot - \omega_N)
\nonumber\\
&= g^2 t 				\int {dk_1\cdot\cdot\cdot dk_N \over {\omega_1\cdot\cdot\cdot \omega_N}}
					\delta(k - k_1 - k_2 + \cdot\cdot\cdot - k_N)	\delta (\omega - \omega_1 - \omega_2 - \cdot\cdot\cdot - \omega_N).
\end{align}

In the final expression, the integral simply calculates the volume of $k-$space for the fields $X_i$ for which the modes
multiply to a resonant mode of $X$, but with the unusual volume element $(\omega_1\cdot\cdot\cdot \omega_N)^{-1}$. 
In fact, as is well known, this volume element is exactly right to make the 
integral independent of $k, \omega$! This can be
seen by writing each $dk_i$ integral in manifestly Lorentz-invariant form as
\begin{align}
\label{dkomega}\int {dk_i\over\omega_i} = 2\int dk_i d\omega_i \,\delta(\omega_i^2 - k_i^2 - m_i^2).
\end{align}
A change to the initial $k, \omega$ (i.e., starting with a faster or slower moving packet)
must preserve $\omega^2 = k^2 + m^2$, hence corresponds 
to a Lorentz transformation in $k,\omega$. This may be counteracted by applying the opposite Lorentz
transformation to each $\omega_i,k_i$ throughout the integral, leading to a velocity-independent 
rate of energy emission from the $X$ wave packet into the fields $X_i$. 

\section{Acknowledgements}
 
The author would like to thank Larry Hoffman and Michael Lennek for helpful comments and discussions.

\end{document}